# Indirect precise angular control using four-wave mixing

Wei Zhang[1,2], Dong-Sheng Ding[1,2], Yun-Kun Jiang[3], Bao-Sen Shi[1,2] * and Guang-Can Guo[1,2]

[1]Key Laboratory of Quantum Information, CAS, University of Science and Technology of China, Hefei, Anhui, 230026, China

[2]Synergetic Innovation Center of Quantum Information & Quantum Physics, University of Science and Technology of China, Hefei, Anhui, 230026, China

[3]Fuzhou University, Fuzhou, Fujian, 350108, China

Here we show indirect precise angular control using a four-wave mixing (FWM) process. This was performed with a superposition of light with orbital angular momentum (OAM) in an M-Type configuration of a hot $^{85}$Rb atomic ensemble. A gear-shaped interference pattern is observed at FWM light with a donut-shaped input signal. The gear could be rotated and is controlled through the change of the polarization of the pump laser. Our experimental results that are based on nonlinear coherent interactions have applications in image processing and precise angular control.

It is generally accepted that light can carry angular momentum including the spin angular momentum (SAM) associated with left or right circular polarization and also the orbital angular momentum (OAM) characterized by the helical phase structure[1]. Although SAM can have only two values, $\pm\hbar$, the OAM that light carry is given by $l\hbar$, where $l$ is an arbitrary large positive or negative integer. SAM is always thought of as the basis of polarization. Light carrying SAM demonstrates the importance of the effect of polarization on atom-field interactions when considering the magnetic sublevels[2,3]. The OAM of photons has attracted interest because of its wide use in imaging[4-6], remote sensing[7], in precise measurements[8] and in quantum processes[9,10].

*Corresponding Author: drshi@ustc.edu.cn

Recently in quantum information and quantum optics, light was encoded with information in its OAM degrees of freedom enabling networks to carry more information and increase their capacity[11, 12]. The OAM of light also has applications in both classical and quantum fields[13]. Recently, Lavery et al. successfully detected the angular frequency of a spinning object using the OAM of light by analyzing the frequency shifts[14]. Also, the quantum entanglement of OAM made by Fickler et al. demonstrated the advances in the improvement of the sensitivity in angular resolution[15]. However, all of these experiments were accomplished using the direct modulation of light or single photons.

Here we show indirect precise angular control by using a nonlinear four-wave mixing (FWM) process with an input laser beam consisting of the superposition of OAM in an M-Type configuration of a $^{85}$Rb atomic ensemble. The two planar counter-propagating pump waves interact with a probe field, yielding a fourth output wave. G. Walker el al. observed that the phase profile associated with OAM is transferred entirely from the pump laser to the generated light through a FWM process in a cascade transition[16]. In our FWM process, we observed a gear-shaped interference pattern of FWM light with a donut-shaped input using the effect of polarization. The two pump lasers are a polarized Gaussian beam compared with the work of G. Walker, which used pump laser beams carrying OAM. Moreover the gear-shaped interference pattern could be rotated when we changed the polarization of one of the pump laser beams. The nonlinear process we used is advantageous not only because it is an indirect method, but also because the wavelength of the generated light can be tuned. Here the wavelength of the FWM was 795 nm, which can be customized in a wide spectrum from the ultraviolet to the infrared through different energy levels[17, 18]. Since the precise control of a physical quantity is a problem in many research areas, we believe our indirect method is useful to many fields.

A diagram of the energy levels is shown in Fig. 1, which consists of four states |1>, |2>, |3> and |4>. The four states |1>, |2>, |3> and |4> correspond to the $^{85}$Rb atomic levels $5S_{1/2}$(F=3), $5S_{1/2}$(F=3), $5P_{3/2}$(F'=4) and $5P_{1/2}$(F''=2). Note that |1> and |2> are degenerate states. The natural line widths of |3> and |4> were approximately 6.1 and 5.8 MHz respectively. The pump 1 laser at 780 nm (DL100, Toptica) was red detuned with the atomic transition |1>→|3>. The transitions |2>→|4> and |4>→|1> were resonantly coupled by the pump 2 and signal laser beams at a wavelength of 795 nm. The pump 2 and signal beams were from the same laser (DL100, Toptica). A 5-cm long $^{85}$Rb cell was heated to 120 ℃ where the density of $^{85}$Rb atom is estimated to be $2\times10^{19}/m^3$.

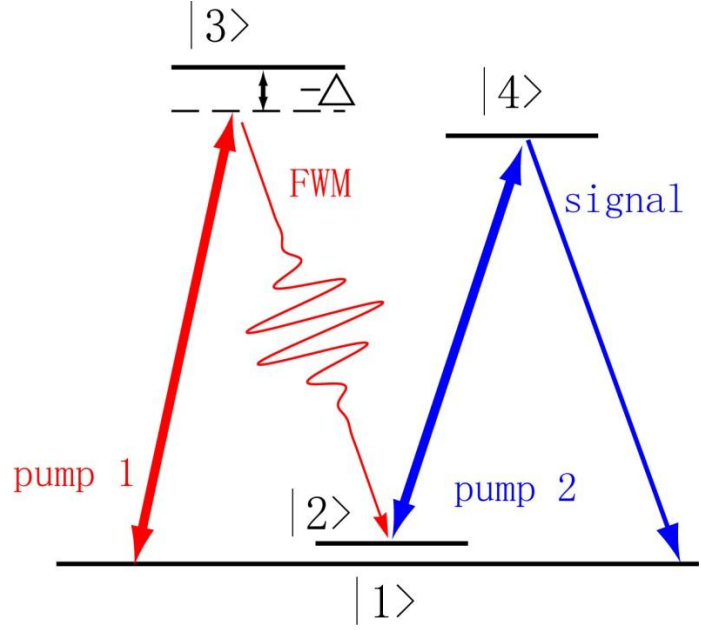

Fig. 1. A diagram of the energy levels used in the experiment. |1> and |2> are degenerate states that correspond to the atomic level $5S_{1/2}$(F=3). |3> and |4> are the corresponding states of atomic levels $5P_{3/2}$(F'=4), $5P_{1/2}$(F''=2). The detuning of pump 1 laser, $\Delta$, is ~-600 MHz. The pump 2 and signal laser are resonant with the atomic transitions.

In our experimental setup, the two pump laser beams propagated nearly collinearly through the $^{85}$Rb cell. There was an angle of 2° between signal beam and the pump beams (see Fig. 2). The two pump laser beams were parallel Gaussian beams. The signal beam carrying OAM was focused to the center of the $^{85}$Rb vapor by a lens with a focal length of ~500 mm. By using a quarter-wave plate (QWP), a half-wave plate (HWP) and a polarization beam splitter (PBS), we controlled the power of the pump 1 and pump 2 beams. The pump 1 beam had horizontal linear polarization, while the polarization of pump 2 was controlled by a HWP (in position C in Fig. 2) with a rotation angle of $\theta$ from vertical axis.

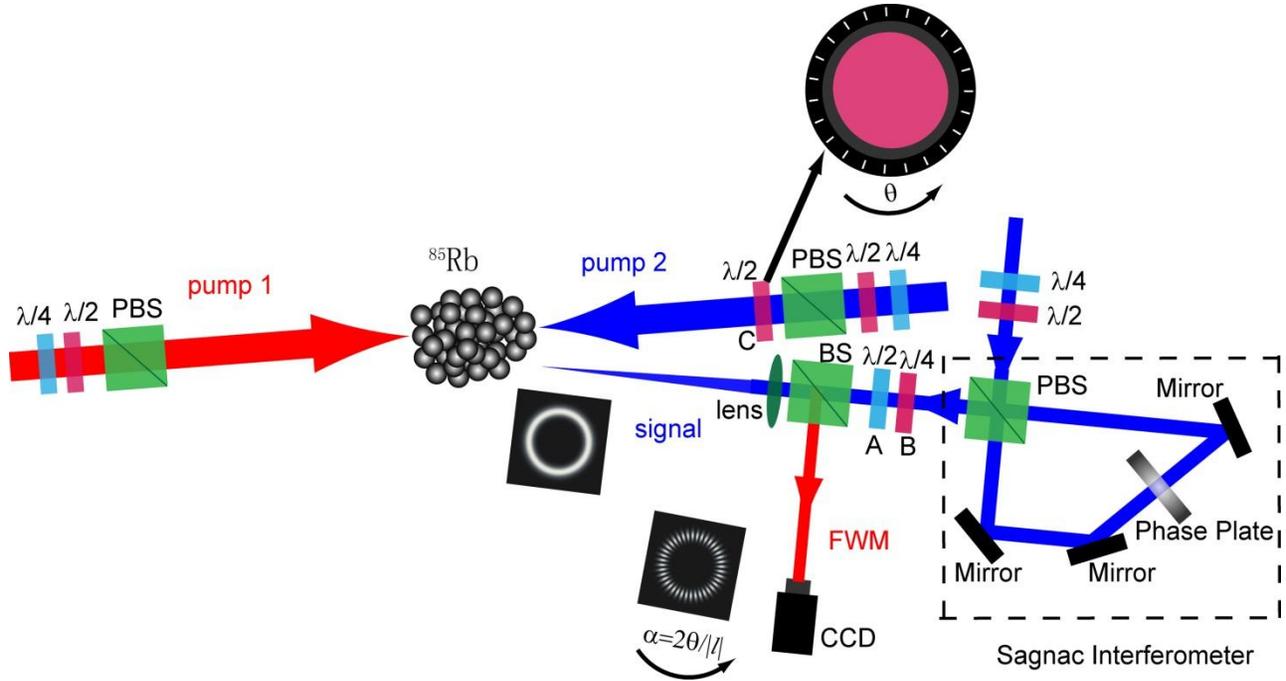

Fig. 2. The experimental setup of the FWM process using different OAM inputs. λ/2 is the half-wave plate and λ/4 is the quarter-wave plate. A, B and C represent the positions of the corresponding elements. PBS is the polarization beam splitter and BS is the beam splitter. The Sagnac Interferometer is the part surrounded by the dashed line including the mirrors, the PBS and the phase plate. The FWM is the generated four-wave mixing light and the CCD is the charge-coupled camera. The angle $\theta$ is the amount that the HWP is rotated in position C. The rotated angular of the interference spots at FWM light is $\alpha$ and $l$ is the quantity of the OAM. The power of the pump 1, pump 2 and the signal laser beams were approximately 9, 9 and 5 $m$W respectively. The focal length of the signal for $|l|=2$ is 540 and 440 mm for $|l|=20$.

Before performing the angular control of the generated FWM light, the input signal light was encoded with OAM using a Sagnac interferometer. The key element in the Sagnac interferometer is the phase plate that makes the horizontal and vertical light carry same OAM with opposite sign. To get a better interference of the beams, we placed the phase plate in the middle of the arm of the interferometer and also made the intensity of the output |H> and |V> polarization equal using a QWP and a HWP before the interferometer. The output state of the signal beam obtained after the interferometer is represented as $|\psi_1\rangle = (|l\rangle|H\rangle + |-l\rangle|V\rangle)$, where |H> and |V> represent the horizontal linear polarization and vertical linear polarization states respectively and |l> is the quantity of the OAM. By inserting a QWP

in position B and a HWP in position A, the state of signal can be represented by

$$|\psi_{input}> = \frac{1}{2}(ie^{i2\theta_0}|l> + e^{-i2\theta_0}|-l>)|H> - \frac{1}{2}(e^{i2\theta_0}|l> + ie^{-i2\theta_0}|-l>)|V>. \tag{1}$$

The fast axis of the QWP had a $\pi/4$ rotation from vertical axis and the fast axis of the HWP was at an angle $\theta_0$ from vertical axis.

If the polarization of pump 2 is $|P_\theta>$ and the polarization perpendicular to the polarization of pump 2 is $|P_\gamma>$, we can rewrite the signal light based on $|P_\theta>$ and $|P_\gamma>$ as follows:

$$|\psi_{input}> = \frac{1}{2}(ie^{-i2(\theta-\theta_0)}|l> + e^{i2(\theta-\theta_0)}|-l>)|P_\theta> + \frac{1}{2}(e^{-i2(\theta-\theta_0)}|l> + ie^{i2(\theta-\theta_0)}|-l>)|P_\gamma>. \tag{2}$$

In our experiment, the efficiency of FWM process is different with different linearly polarized pump and signal lasers. This is mainly because of the different interaction intensities in the different situations. If a linear polarized pump 2 laser beam pumps the atoms from |2> to |4> with the condition of $\Delta m_F = 0$ where $m_F$ is the magnetic quantum number, then the same linearly polarized weak signal can pump the atoms from |4> back to |1> with $\Delta m_F = 0$. The whole process may be seen as a $\pi$-$\pi$ transition process in which $\pi$ represents the linear polarized light. But if the signal light changed to another polarization (for example, vertical to the linear polarization of pump 2), this must be taken to be a superposition of left-circular polarization and right-circular polarization accompanied with $\Delta m_F = \pm 1$ in the transition. The whole process could be seen as $\pi$-$\sigma^\pm$ transition process [2], which will result in the accumulation of a population in the uncoupled Zeeman sublevels with $m_F = \pm 3$. The $\pi$-$\pi$ process has a stronger interaction than $\pi$-$\sigma^\pm$ process. In our FWM generation process, the same effect of polarization has great influence. Also the situation is more complicated as it is related to four energy levels. With the conditions of phase-matching and the conversion of energy, the output state monitored by using a BS and a CCD is

$$|\psi_{FWM}> \sim \beta(ie^{-i2(\theta-\theta_0)}|l> + e^{i2(\theta-\theta_0)}|-l>)|H> + (e^{-i2(\theta-\theta_0)}|l> + ie^{i2(\theta-\theta_0)}|-l>)|V>, \tag{3}$$

where the scale factor coefficient $\beta$ results from the effect of the polarization in the transition process. In the practical detection process, the noise from the FWM of the vertical linear polarization is less than the FWM of horizontal linear polarization by a factor of 4.5, which corresponds to $\beta=2.1$. So we could observe the interference pattern. In this case, the detected FWM is

$$|\psi_{FWM}>\sim \frac{1}{\sqrt{2}}(ie^{-i2(\theta-\theta_0)}|l>+e^{i2(\theta-\theta_0)}|-l>)|H>. \tag{4}$$

This result corresponds to the gear-shaped interference spots that were observed in our experiment and are shown in Fig. 3. Fig. 3(a), (b), (c) and (d) are the theoretical simulations and Fig. 3(e), (f), (g) and (h) are the experimental results. This is for the case where $|l|=2$ and $|l|=20$.

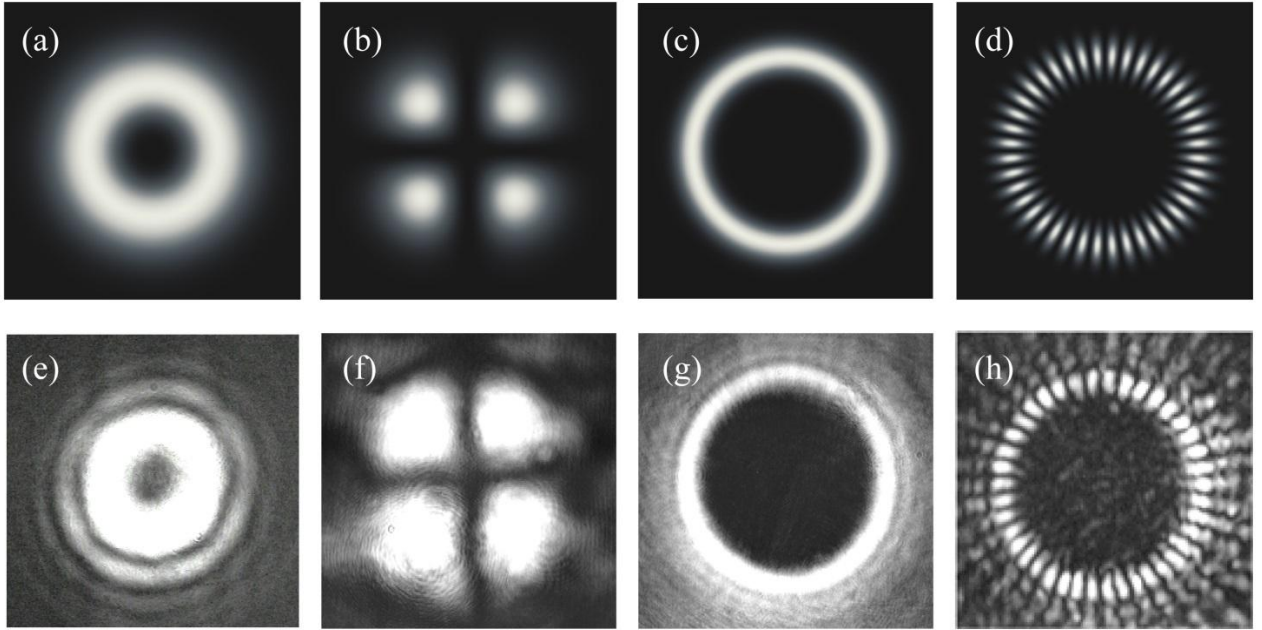

Fig. 3. Intensity distribution corresponding to the signal and the FWM of the light. (a) and (b) are theoretical intensity distributions that correspond to the signal and FWM for $|l|=2$. (c) and (d) are theoretical intensity distributions that correspond to the signal and FWM for $|l|=20$. (e), (f), (g) and (h) are the experimental intensity distributions that correspond to (a), (b), (c) and (d), respectively.

Experimentally, when we rotate the HWP in position C that has the fast axis an angle $\theta$ from the vertical axis, we observed a rotation of $\alpha$ of the interference spots at FWM accordingly. From Eq. (4), we know that the angular cycle of $\theta$ is $T_\theta = 2\pi/4 = \pi/2$, and the cycle of the interference spots is $T = 2\pi/(2|l|) = \pi/|l|$. The angular ratio is

$$\alpha = \frac{T}{T_\theta}\theta = \frac{2}{|l|}\theta. \tag{5}$$

We can see that a rotation of α has a linear relationship with the rotation of θ scaled by $2/|l|$. Apparently, a large value of *l* can be obtained using the phase plate with high mode number of OAM or using a spatial light modulator (SLM), and so we can control the angular rotation α precisely by rotating the HWP. Here in Fig. 4 we use $|l|=2$ to illustrate the generated FWM with the change of θ both theoretically and experimentally.

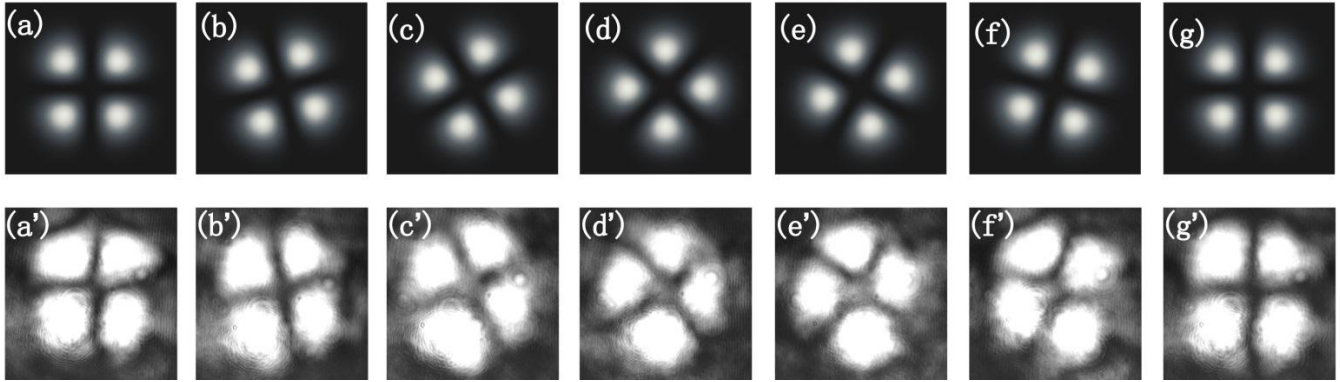

Fig. 4. The generated FWM light with the change of θ. The theoretical simulations are shown in (a), (b), (c), (d), (e), (f) and (g) with θ = 0°, 15°, 30°, 45°, 60°, 75° and 90° respectively. The corresponding experimental results taken with a CCD camera are shown in (a′), (b′), (c′), (d′), (e′), (f′) and (g′) respectively.

Clearly, we can see the rotation of the FWM with the change of the angle of θ coming from the rotation of the HWP. To verify Eq. (5), we recorded the change of the rotation. The rotation by the angle of θ was read from the scale on the HWP, and the rotation by α was measured using the photos taken with the CCD camera. Here, when we measured the angle of α, we located a cross line in the center of the photo connecting to center of the four selected spots. The rotation of the gear-shaped interference spots was measured using the rotation of the cross line to get a more accurate result. We can clearly see that the experimental data fit with theoretical predictions as shown in Fig. 5.

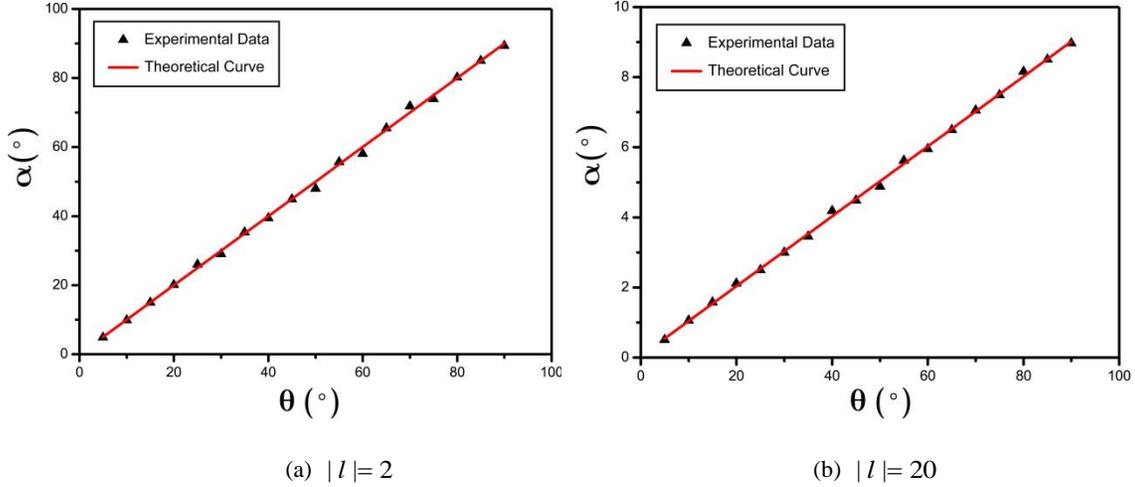

(a) $|l|=2$                                        (b) $|l|=20$

Fig. 5. The relation between the rotation of angle $\alpha$ and the rotation of angle $\theta$: (a) The relationship between the rotation $\alpha$ and the rotation $\theta$ for $|l|=2$; (b) the relationship between the rotation $\alpha$ and the rotation $\theta$ for $|l|=20$.

Here we show indirect precise angular control through a nonlinear process and the analysis of the physical principles behind it. The nonlinear process is of significance because of its wide spectrum. Since D'Ambrosio et al. proposed the concept of "photonic polarization gears" for ultra-sensitive angular measurements by considering the rotation amplification[13], our work may be seen as a "micrometer" if the "scaling-down" effect of the rotation is taken into consideration. In Eq. (4), we can see that the rotation can be controlled by $\theta_0$ and $\theta$ simultaneously, which was also shown during our experiments. We can therefore extend our indirect angular controller to a double angular controller that may be used in precise optical switching. We believe our method is useful for research in the field of OAM and for the precise control of light.

This work was supported by the National Natural Science Foundation of China (Grant numbers 11174271, 61275115 and 10874171), the National Fundamental Research Program of China (Grant number 2011CB00200), the Youth Innovation Fund from USTC (Grant number ZC 9850320804), the Innovation Fund from CAS, Program for NCET.